\documentclass[prd,twocolumn,superscriptaddress,showpacs,nofootinbib,
preprintnumbers]{revtex4}

\usepackage{amsmath}
\usepackage{amsfonts}
\usepackage{graphicx}
\usepackage{dcolumn}

\def\be{\begin{equation}}
\def\ee{\end{equation}}
\def\ba{\begin{eqnarray}}
\def\ea{\end{eqnarray}}
\def\bs{\begin{subequations}}
\def\es{\end{subequations}}

\usepackage{color}


\begin{document}

\title{The thawing dark energy dynamics: Can we detect it? }

\author{S. Sen}
\affiliation{Centre for Theoretical Physics, Jamia Millia Islamia,
New Delhi-110025, India}
\author{A. A. Sen}
\affiliation{Centre for Theoretical Physics, Jamia Millia Islamia,
New Delhi-110025, India}
\author{M. Sami}
\affiliation{Centre for Theoretical Physics, Jamia Millia Islamia,
New Delhi-110025, India}

\begin{abstract}
We consider different classes of scalar field models including
quintessence and tachyon scalar fields with a variety of generic
potential belonging to the thawing type. We focus on observational
quantities like Hubble parameter, luminosity
 distance as well as quantities related to the Baryon Acoustic Oscillation
 measurement. Our study shows that with present state of observations, one
 can not distinguish amongst various models which in turn can not be
 distinguished from cosmological constant. Our analysis indicates 
that there is a thin chance to observe the dark energy metamorphosis in near
 future.
\end{abstract}
\maketitle

\section{Introduction}
The fact that our universe is currently going through an accelerated
phase of expansion is one of the most significant discoveries
\cite{sndis} in physics in recent times that can have far reaching
implications for fundamental theories of physics. Late time cosmic
 acceleration can
be fueled either by assuming the presence of an exotic fluid with
large negative pressure known as {\it dark energy} or by modifying
gravity itself. The simplest candidate of dark energy is provided by
cosmological constant with equation of state parameter $w=-1$.
However, the model based upon cosmological constant is plagued with
the fine tuning and cosmic coincidence problems (See Ref \cite{rev}
 for a nice review).

Scalar field models with generic features can alleviate the fine
tuning and coincidence problems and provide an interesting
alternative to cosmological constant \cite{scalar}. The simplest
generalization of cosmological constant is provided by a scalar
field with linear potential \cite{linear}. Its evolution begins from
the locking regime (due to large Hubble damping) where it mimics the
cosmological constant like behavior. At late times, the field starts
rolling and since the potential has no minimum, the model leads to a
collapsing universe with a finite history.

The more complicated scalar field models can broadly be classified
into two categories- fast roll and slow roll models dubbed freezing
and thawing models \cite{thaw}.  In case of the fast roll models,
the potential is steep  allowing the scalar field to mimic the
background being subdominant for most of the evolution history. Only
at late times, the field becomes dominant and drives the
acceleration of the universe. Such solutions are referred to as {\it
trackers}.

Slow-roll models are those for which the field kinetic energy is
much smaller than its potential energy. It usually has a
sufficiently flat potential similar to an inflaton. At early times,
the field is nearly frozen at $w=-1$ due to the large Hubble
damping. Its energy density is nearly constant and and its
contribution to the total energy density of the universe is also
nearly negligible. But as radiation/matter rapidly dilutes due to
the expansion of the universe and the background energy density
becomes comparable to field energy density, the field breaks away
from its frozen state evolving slowly to the region with larger
values of equation of state parameter. In this case, however, one
needs to have some degree of fine tuning of the initial conditions
in order to achieve a viable late time evolution.

Recent observations suggest that the equation of state parameter for
dark energy
 does not significantly deviate from $w=-1$ around the present epoch
\cite{wood}.
   This type of equation of state can be easily obtained in dynamical
models represented by thawing scalar fields. This fact was exploited in
Ref \cite{scsen1} which examined quintessence models
  with nearly flat potentials satisfying the slow-roll conditions. It was
shown that under the slow-roll conditions,  a scalar field with a variety
of potentials $V(\phi)$ evolve in a similar fashion and one can derive a
generic expression for equation of state for all
  such scalar fields. This result was later extended to the case of phantom \cite{scsen2}
  and tachyon scalar fields \cite{amna}. It was demonstrated that under
  slow-roll conditions, all of them have identical equation of state and hence can
  not be distinguished, at least, at the level of background cosmology. The crucial assumption, for
  arriving  at this important conclusion was the fulfillment of the slow-roll condition for the field potentials.

In this paper, we relax the assumption of slow roll but assume that
the scalar field is of thawing type i.e it is initially frozen at
$w=-1$ due to large Hubble damping. With non-negligible matter
contribution, this does not necessarily mean the small value for
$V_{,\phi}/{V}$ which is the usual slow-roll parameter for inflaton.
We, rather, assume that the slow-roll condition is highly broken
such that
 $V_{,\phi}/{V} \sim 1$. In this case, we need to fine tune the initial conditions
 to match the observational value of the present day dark energy density which is a characteristic feature
 of any thawing model. With this choices, we study the evolution
 of a variety of scalar field models having both canonical and non-canonical kinetic terms.
 We particularly focus on the observational quantities like Hubble parameter, luminosity
 distance as well as quantities related to the Baryon Acoustic Oscillation (BAO) measurement.

\section{Thawing Scalar Field}

In what follows, we shall assume that the dark energy is described
by a minimally-coupled scalar field, $\phi$, with equation of motion
\begin{equation}
\ddot{\phi}+3 H{\phi}+ V_{,\phi}=0 \label{motionq}
\end{equation}
where the Hubble parameter $H$ is given by
\begin{equation}
\label{H}
H = \left(\frac{\dot{a}}{a}\right) = \sqrt{\rho/3}.
\end{equation}
Here $\rho$ is the total energy density in the universe. We model a
flat universe containing only matter and a scalar field, so that
$\Omega_\phi + \Omega_M = 1$.

Equation (\ref{motionq}) indicates that the field rolls downhill in
the potential $V(\phi)$, but its motion is damped by a term
proportional to $H$. The equation of state parameter $w$ is given by
$w=p_{\phi}/\rho_{\phi}$ where the pressure and density of the
scalar field have the form
\begin{eqnarray}
&&p = \frac{\dot \phi^2}{2} - V(\phi) ,\\
&&\rho = \frac{\dot \phi^2}{2} + V(\phi)
\end{eqnarray}
Observations suggest a value of $w$ near $-1$ around the present
epoch. We adopt a similar technique as followed in references
\cite{scsen1,scsen2} and define the variables $x$, $y$, and $\lambda$ as
\begin{eqnarray}
\label{xevol}
x &=& \phi^\prime/\sqrt{6}, \\
y &=& \sqrt{V(\phi)/3H^2}, \\
\lambda &=& -\frac{1}{V}\frac{dV}{d\phi},
\end{eqnarray}
where prime as usual denote the derivative with respect to $\ln a$:
e.g., $\phi^\prime \equiv a(d\phi/da)$

Then contribution of the kinetic energy and potential energy of the
scalar field to the fractional density parameter $\Omega_\phi$ are
represented by $x^2$ and $y^2$ such that,
\begin{equation}
\label{Om}
\Omega_\phi = x^2 + y^2,
\end{equation}
while the equation of state is given by,
\begin{equation}
\label{gamma}
\gamma \equiv 1+w = \frac{2x^2}{x^2 + y^2}.
\end{equation}

In terms of the variables $x$, $y$, and $\lambda$, evolution
equations (\ref{motionq}) and (\ref{H}) take the autonomous form
\begin{eqnarray}
x^\prime &=& -3x + \lambda\sqrt{\frac{3}{2}}y^2 + \frac{3}{2}x[1 + x^2-y^2],\\
y^\prime &=& -\lambda\sqrt{\frac{3}{2}}xy + \frac{3}{2}y[1+x^2-y^2],\\
\lambda^\prime &=& - \sqrt{6} \lambda^2(\Gamma - 1) x,
\end{eqnarray}
where
\begin{equation}
\label{Gamma} \Gamma \equiv V \frac{d^2
V}{d\phi^2}/\left(\frac{dV}{d\phi} \right)^2.
\end{equation}
We now rewrite these equations, changing the dependent variables
from $x$ and $y$ to the observable quantities $\Omega_\phi$ and
$\gamma$ given by equations (\ref{Om}) and (\ref{gamma}). To make
this transformation, we assume that $x^\prime > 0$; our results
generalize trivially to the opposite case. In terms of $\Omega_\phi$
and $\gamma$ the above set of equation become
\begin{eqnarray}
\label{gammaprime}
\gamma^\prime &=& -3\gamma(2-\gamma) + \lambda(2-\gamma)\sqrt{3 \gamma
\Omega_\phi},\\
\label{Omegaprime}
\Omega_\phi^\prime &=& 3(1-\gamma)\Omega_\phi(1-\Omega_\phi),\\
\label{lambda}
\lambda^\prime &=& - \sqrt{3}\lambda^2(\Gamma-1)\sqrt{\gamma \Omega_\phi}.
\end{eqnarray}

This is an autonomous system of equations involving the observable parameter $\gamma$ and $\Omega_{\phi}$.
 Given the initial conditions for $\gamma$, $\Omega_{\phi}$ and $\lambda$, one can solve this system of
 equation numerically for different potentials. As we mention earlier, we are interested in thawing models
 i.e models for which the equation of state is initially frozen at $w=-1$. Hence $\gamma = 0$ initially
 for our purpose. We also do not assume slow-roll conditions for the scalar field potentials rather we
 consider situations for which it is broken strongly i.e $\lambda_{initial} \sim 1$. We should mention
 that for models where slow-roll condition is satisfied i.e. $\lambda <<1$, it has been already shown that all
  such models have an identical equation of state as a function of scale
factor \cite{scsen1, scsen2}. In general the contribution
  of the scalar field to the total energy density of the universe is insignificant at early times, nevertheless
  one has to fine tune the initial value of $\Omega_{\phi}$ in order to have its correct contribution at present.
  This is the fine tuning one needs to have in a thawing models. With these initial conditions we evolve the above
  system of equations from redshift $z=1000$ (or $a = 10^{-3}$) till the present day $z=0$ ($a=1$).
We consider various types of potentials e.g $V=\phi$, ~$V = \phi^2$,
~ $V=e^\phi$ and $V = \phi^{-2}$, characterized by
$\Gamma=0,~\frac{1}{2},~1$ and $\frac{3}{2}$ respectively.
 We have taken
two sets of solution such that $\Omega_{\phi}=0.7$ and
$\Omega_{\phi}=0.75$ at the present epoch for all chosen values of
$\Gamma$.

We also consider the Pseudo-Nambu Goldstone Boson (PNGB) model \cite{pngb}.  (For a recent
discussion, see Ref. \cite{Albrecht} and references therein).
This model is characterized by the potential
\begin{equation}
V(\phi) = M^4 [\cos(\phi/f)+1],
\end{equation}
Alam et al.,\cite{starob} have previously considered such type of potential to see whether dark energy is decaying or not.

We have chosen $f$ to be $1$ for our purpose without any loss generality.

\begin{figure}[t]
\includegraphics[width=80mm]{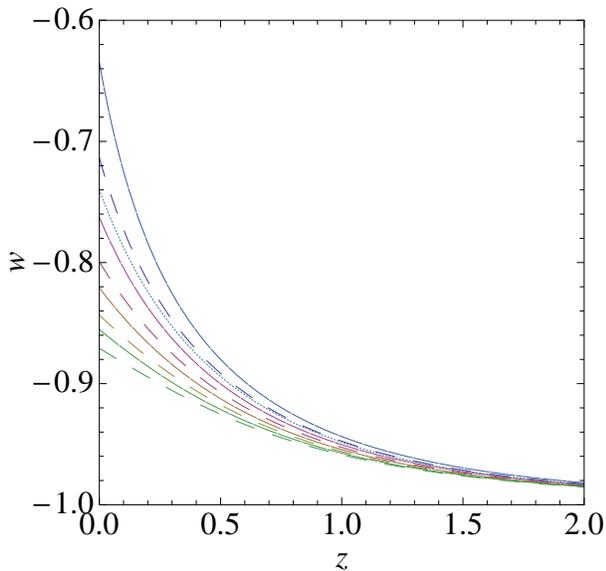}
\caption{Plot of equation of state $w$ vs. redshift for different scalar field
and tachyon models.
Solid curves represent different Tachyonic models with $V(\phi) = \phi, \phi^2, e^{\phi}, \phi^{-2}$ respectively from top to bottom,
Dashed curves from top to bottom represent different scalar field models with same
potentials as in tachyon. Dotted curve represents PNGB model. $\Omega_{m} = 0.3$.}
\end{figure}

As mentioned before, tachyon field is of interest in cosmology.
There have been several investigations using this field as a dark
energy candidate \cite{tachyon}. In what follows, we shall repeat
the above presented analysis  for tachyon field.
\section{Thawing Tachyon Field}

The tachyon field is specified by the Dirac-Born-Infeld (DBI) type
of action \cite{sen}

\begin{equation}
{\mathcal{S}}=\int {-V(\phi)\sqrt{1-\partial^\mu\phi\partial_\mu\phi}}\sqrt{-g} d^4x.
\label{Taction1}
\end{equation}
In FRW background, the pressure and energy density of the tachyon field $\phi$
are given by
\begin{equation}
p_{\phi}=-V(\phi)\sqrt{1-\dot{\phi}^2}
\end{equation}
\begin{equation}
\rho_{\phi}=\frac{V(\phi)}{\sqrt{1-\dot{\phi}^2}}
\end{equation}
The equation of motion which follows from (\ref{Taction1}) is
\begin{equation}
\ddot{\phi}+3H\dot{\phi}(1-\dot{\phi}^2)+\frac{V'}{V}(1-\dot{\phi}^2)=0
\end{equation}
where $H$ is the Hubble parameter.
The evolution equations can be cast in the following autonomous form
for the convenient use
\begin{eqnarray}
&& x'_t=-(1- x_t^2)(3x_t-\sqrt{3}\lambda_t y_t)\\
&& y'_t=\frac{y_t}{2}\left[-\sqrt{3} \lambda_t x_t
y_t-\frac{3(1- x_t^2)y_t^2}{\sqrt{1- x_t^2}}+3\right] \\
&&\lambda'_t=-\sqrt{3}\lambda_t^2 x_t y_t(\Gamma-\frac{3}{2})
\end{eqnarray}
with $x_t,y_t,\lambda_t$ and $\Gamma$ defined as
\begin{eqnarray}
 x_t=\dot{\phi},
 ~y_t=\frac{\sqrt{V(\phi)}}{\sqrt{3}H},~\lambda_t=-\frac{V_{\phi}}{V^{\frac{3}{2}}},~\Gamma=V\frac{V_{\phi\phi}}{V_{\phi}^2}
 \label{Gamma1}
\end{eqnarray}
where prime again denotes the derivative with respect to $\ln(a)$.
\begin{figure}[t]
\includegraphics[width=80mm]{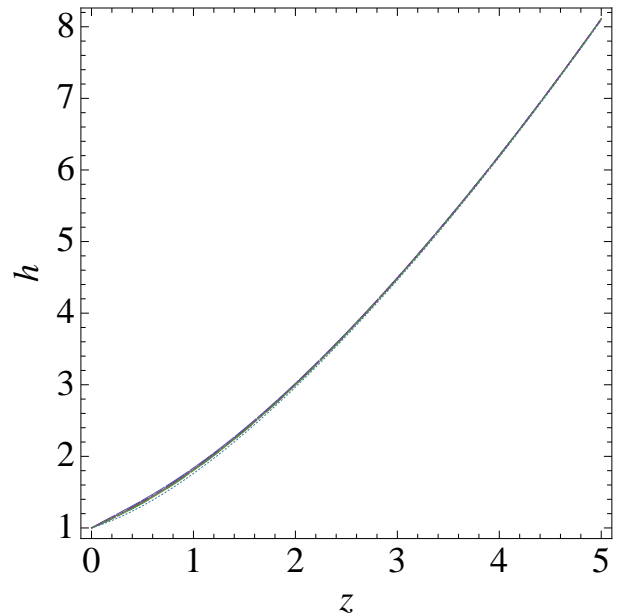}
\caption{Plot of $h(a) = \frac{H(a)}{H_{0}}$ vs. redshift for
different models as well as for $\Lambda$CDM corresponding to
$\Omega_{m0}=0.3$.}
\end{figure}
We would further use the following definitions for the tachyon field
as we did in case of thawing quintessence
\begin{eqnarray}
\Omega_{\phi}=\frac{y^2}{\sqrt{1-x^2}},~~\gamma \equiv 1+w=\dot{\phi}^2,
\end{eqnarray}
where $w= {p_{\phi}\over{\rho_{\phi}}}$ is the equation of state for the tachyon field. One can
now express the autonomous equations through them:
\begin{eqnarray}
\gamma'=-6\gamma(1-\gamma)+2\sqrt{3\gamma\Omega_{\phi}}\lambda_t(1-\gamma)^\frac{5}{4}
\label{tachgamma}\\
\Omega_{\phi}'=3\Omega_{\phi}(1-\gamma)(1-\Omega_{\phi})
\label{tachomega}\\
\lambda_t'=-\sqrt{3\gamma\Omega_{\phi}}\lambda^2(1-\gamma)^\frac{1}{4}(\Gamma-\frac{3}{2})
\label{tachlambda}
\end{eqnarray}

 We adopt a similar treatment to solve
 the above set of equation (\ref{tachgamma})-(\ref{tachlambda}) as we had
done in the earlier case. Infact, we even consider similar kind of
potentials for tachyon fields as well, i.e, $V=\phi$,~$V = \phi^2$,~
$V=e^\phi$ and $V = \phi^{-2}$, characterized by
$\Gamma=0,~\frac{1}{2},~1$ and $\frac{3}{2}$ respectively along with
the initial conditions for $\lambda_{t initial}$ to be $1$ and
$\gamma_{initial}\sim 0$.  Here also we take two  solutions set for
all $\Gamma$'s, with two different initial conditions of
$\Omega_{\phi}$ such that at present it contributes $70\%$ and
$75\%$ of the total energy share. Before discussing our result, we
want to point that the system of equations (14)-(16) and (27)-(29)
for scalar field and tachyon field respectively are completely
different. Hence {\it a priori} one expects to have different
evolutions for different potentials as well as for scalar and
tachyon fields.


Once we know the solution for $\Omega_{\phi}(a)$ by solving either
(14)-(16) or (27)-(29), we can easily find the behavior of the
Hubble parameter which, in terms of $\Omega_{\phi}$, can be
expressed as
\begin{equation}
h^2(a)=\frac{H^2(a)}{H_0^2}=\frac{1-\Omega_{\phi 0}}{1-\Omega_{\phi}} a^{-3},
\label{hz}
\end{equation}
where $H_{0}$ and $\Omega_{\phi0}$ are the present day values for
the Hubble parameter and the dark energy density parameter. This is
the most important parameter as all the observable quantities
involving background cosmology can be constructed from this.
Moreover there are independent observational constraint on this
parameter itself. We should mention that in this approach, one does
not need to know the equation of state ($\gamma(a) = 1+w(a)$) to
construct the observational quantities although its effect comes
through the solutions of Eqs.(14)-(16) or (27)-(29).
\begin{figure}[t]
\includegraphics[width=80mm]{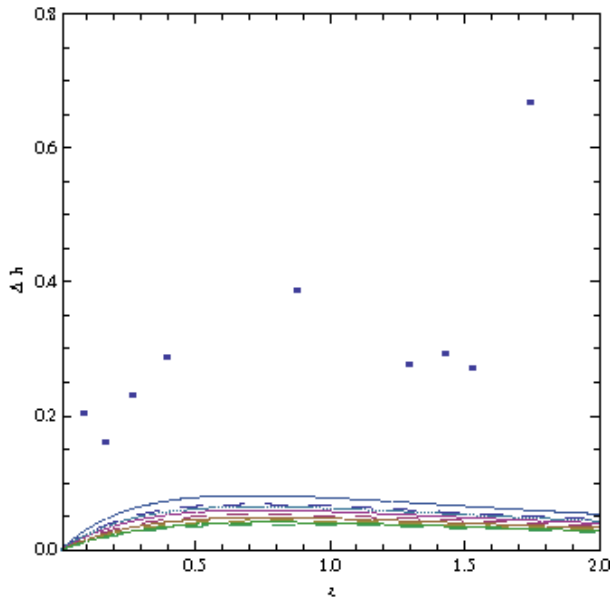}
\caption{Plot of $\Delta h$(explained in the text) vs. redshift for different models. Solid curves represent different Tachyonic models with $V(\phi) = \phi, \phi^2, e^{\phi}, \phi^{-2}$ respectively from top to bottom,
Dashed curves from top to bottom represent different scalar field models with same potentials as in tachyon. Dotted curve represents PNGB model. $\Omega_{m0}=0.3$. The black dots represents the value of the error bar as quoted in \cite{hubble, ratra}. We assume $h_{0} = 73 \pm 3 km s^{-1} Mpc^{-1}$ for our purpose.}
\end{figure}

\section{Results}

Let us now discuss the results of our investigations. In Fig1, we
plot the behavior of the equation of state $w(a)$ for different
models. It shows that the equation of states of different fields
with different potentials behave differently as one approaches the
present day although in the past their behavior are almost identical
which is not surprising as we have assumed the violation of
slow-roll condition, i.e $\lambda_{initial} \sim 1$. With slow-roll
condition satisfied,i.e, $\lambda_{i} << 1$, it was shown earlier
that models with different potentials have the identical $w(a)$ both
for scalar and tachyon fields \cite{scsen1,scsen2,amna}.
\begin{figure}[t]
\includegraphics[width=80mm]{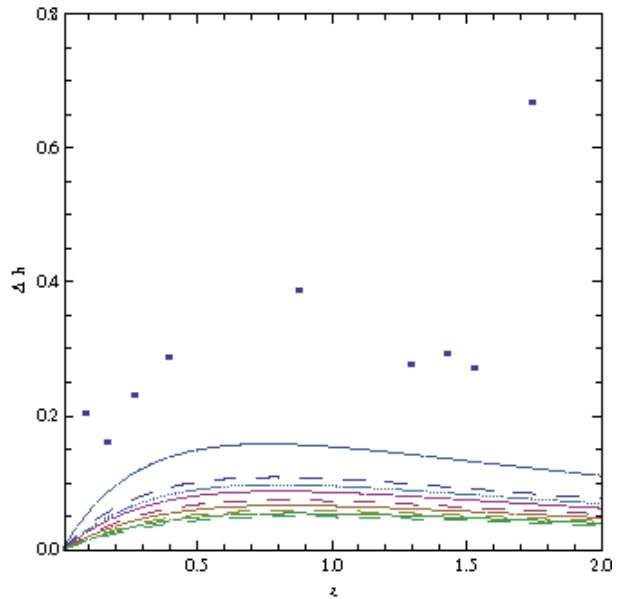}
\caption {Plot of $\Delta h$ vs. redshift for different models same
as Fig 3 but with $\Omega_{m0}=0.25$.}
\end{figure}

Next we investigated the behavior of the Hubble parameter in
different cases. With different behavior for $w(a)$ as shown in
Fig1, one would expect to have different behavior for Hubble
parameter also. However, they are completely indistinguishable as
shown in Fig 2. In this figure, we have also plotted $h(a)$ for
$\Lambda$CDM. It is seen that we can not differentiate  between
individual models as well as these models from $\Lambda$CDM. It is
interesting to note that despite having completely different
behavior for equation of state, all the models have almost identical
 evolution for the Hubble parameter. This is crucial as all the observational quantities are constructed out of $h(a)$ at
 least at the level of background cosmology. In recent past, estimates of $H(z)$ were derived by Simon, Verde and Jimenez
 using passively evolving galaxies \cite{hubble} (also see Ref\cite{ratra}).  Keeping this in mind, we next plot
 the different $\Delta h = h_{field} - h_{\Lambda CDM}$ for each of our
 model in Fig. 3 and Fig. 4 for two different values of $\Omega_{m0}$ ($0.3$ and $ 0.25$) respectively.
In the same figures, we have also plotted the values of the error bars $\Delta h$
(where $\Delta h = {\Delta H\over{H_{0}}} + {H\Delta
H_{0}\over{H_{0}^{2}}}$). We
take the values of $H$ and $\Delta H$ from the the data \cite{hubble, ratra} and use the prior $H_{0} =
73 \pm 3 km s^{-1} Mpc^{-1}$ as quoted therein. Fig.3 $\&$4 show
that with $\Omega_{m0}=0.3$, one can not distinguish all the models
from $\Lambda$CDM as the difference between them is much smaller
than the present error bars. With smaller
  values of the density parameter, $\Omega_{m0} = 0.25$, the difference becomes larger but the error bars
   still
  do not allow  to distinguish the models from $\Lambda$CDM. The
    other interesting feature which one notices from both these figures, is that the difference is maximum in
    the redshift range from $z=0.5$  to $z=1$. Hence having more data points for higher redshifts may not be useful
    for the purpose
   to distinguish different models from $\Lambda$CDM; rather the low redshift measurements is more vital
    to do the needful.
\begin{figure}[t]
\includegraphics[width=80mm]{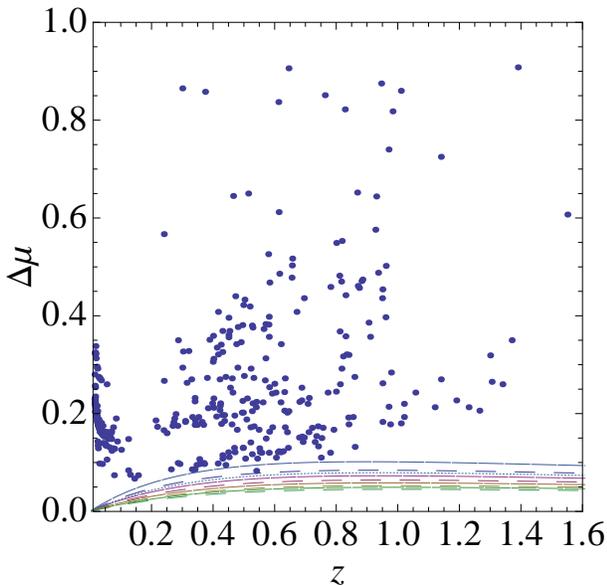}
\caption{Plot of $\Delta \mu$(explained in the text) vs. redshift for different models. Solid curves represent different Tachyonic models with $V(\phi) = \phi, \phi^2, e^{\phi}, \phi^{-2}$ respectively from top to bottom,
Dashed curves from top to bottom represent different scalar field models with same potentials as in tachyon. Dotted curve represents PNGB model. $\Omega_{m0}=0.3$. The black dots represent the value of the error bars as quoted in the Constitution data set \cite{cons}.}
\end{figure}
\begin{figure}[t]
\includegraphics[width=80mm]{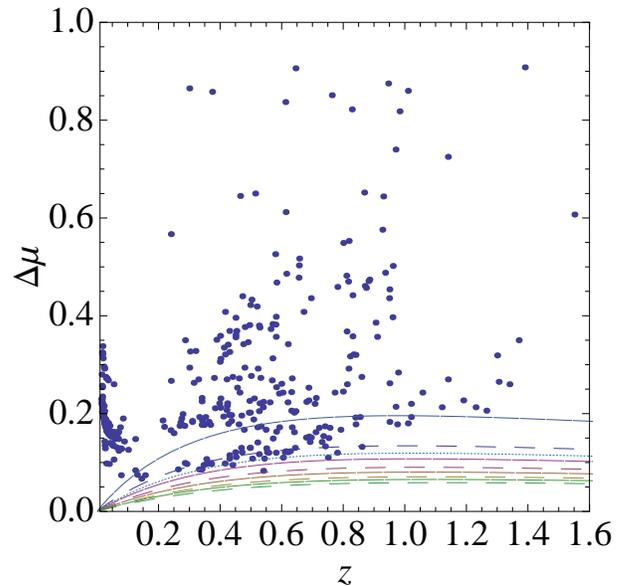}
\caption{ Plot of $\Delta \mu$ vs. redshift for different models
same as Fig 5 corresponding to  $\Omega_{m0}=0.25$.}
\end{figure}

 Next we consider the Supernova Type-Ia observation which is one of the direct
 probes for late time acceleration. It measures the apparent
 brightness of the supernovae as observed by us which is related to the luminosity
 distance $d_{L}(z)$ defined as
\begin{equation}
d_L=(1+z) \int_{0}^{z} \frac{dz^\prime}{H(z^\prime)}.
\label{dl}
\end{equation}
$$
\begin{array}{|c|c|c|}
\hline
 Model & V(\phi) & D_{V\phi}-D_{V\Lambda}\\
  & &for~ \Omega_\phi=0.70  \\
\hline
Thawing Model & \phi~ & -0.01464 \\
\hline
Thawing~model & \phi^2  & -0.01156 \\
\hline
Thawing~model & e^\phi &  -0.00972 \\
\hline
Thawing~model & \frac{1}{\phi^2}  & -0.00843\\
\hline
Scalar~field &  M^4 [\cos \phi+1]~(PNGB) & -0.01398\\
\hline
Tachyon & \phi & -0.01715 \\
\hline
Tachyon & \phi^2 & -0.01291\\
\hline
Tachyon & e^\phi & -0.01062\\
\hline
Tachyon & \frac{1}{\phi^2} & -0.00911\\
\hline
\end{array}
$$
{\centerline{\em Table 1}}
\vspace{2mm}

With this, one constructs the distance modulus $\mu$ which is
experimentally measured:
\begin{equation}
\mu= m - M = 5log\frac{d_L}{Mpc}+25,
\label{dm}
\end{equation}
where $m$ and $M$ are the apparent and absolute magnitudes of the
supernovae which are logarithmic measure of flux and luminosity
respectively. In Fig 5 and Fig 6, we plot the difference $\Delta \mu
= \mu_\phi-\mu_{\Lambda CDM}$ for each model together with current
error bars as quoted in the latest Constitution data set
\cite{cons}. One can now see that with $\Omega_{m0} = 0.3$ (Fig 5),
the difference with $\Lambda$CDM for any model is quite small as
compared to the value of the error bars.  But with $\Omega_{m0} =
0.25$, this difference enhances, and models like tachyon and scalar
field with linear potentials as well as scalar field with PNGB
potential, have significant difference with $\Lambda$CDM which is in
the range of the values of the error bars. The plots also shows that
the intermediate redhsift range between 0.4 and 1.0 is most
sensitive to compare the models with $\Lambda$CDM  and future
experiments involving Type-Ia supernova should focus more in this
redshift range in order to investigate the nature of dark energy.

$$
\begin{array}{|c|c|c|}
\hline
 Model & V(\phi) & D_{V\phi}-D_{V\Lambda}\\
 & &for~ \Omega_\phi=0.75  \\
\hline
Thawing Model & \phi~ & -0.02369 \\
\hline
Thawing~model & \phi^2  & -0.01667\\
\hline
Thawing~model & e^\phi  & -0.01332 \\
\hline
Thawing~model & \frac{1}{\phi^2}   & -0.01121\\
\hline
Scalar~field &  M^4 [\cos \phi+1]~(PNGB) & -0.02156\\
\hline
Tachyon & \phi  & -0.03313 \\
\hline
Tachyon & \phi^2 & -0.01951 \\
\hline
Tachyon & e^\phi  & -0.01498 \\
\hline
Tachyon & \frac{1}{\phi^2}  & -0.01235\\
\hline
\end{array}
$$
{\centerline{\em Table 2}}

\vspace{2mm} Another observational probe that has been widely used
in recent times to constraint dark energy models is related to the
data from the Baryon Acoustic Oscillations measurements\cite{sdss}.
In this case, one needs to calculate the parameter $D_{V}$ which is
related to the angular diameter distance as follows
\begin{equation}
D_V(z_{BAO})=\left[\frac{z_{BAO}}{H(z_{BAO})}(\int_0^{z_{BAO}}\frac{dz}{H(z)})^2\right]^{1/3}
\label{bao}
\end{equation}

For BAO measurements we calculate the ratio
$D_V(z=0.35)/D_V(z=0.20)$. As shown in \cite{percival} this ratio is
a relatively model independent quantity and has a value $1.812 \pm
0.060$. For our case, we calculate the difference of this ratio
between any scalar field model and $\Lambda$CDM model. In tables 1
$\&$ 2, we quote our result for two different values for
$\Omega_{m0}: \Omega_{m} = 0.3$, $\Omega_{m0} = 0.25$ .

As one can see from the results quoted in these two tables, with
current BAO measurements, it is hard to distinguish all the models
from $\Lambda$CDM. One has to decrease the error bars at least by
fifty percent for this purpose.

\section{Conclusion}
In this paper, we have studied the general classes of thawing models
with both quintessence
 and tachyon type scalar fields without assuming slow-roll conditions for the
 potentials of these fields.
 Our investigations show that the overall Hubble parameter has almost
 identical behavior for all these models and also matches with its counterpart
 corresponding to
  $\Lambda$CDM despite of the fact that the equation of state for
 different models behaves quite differently. Since all the observable quantities related to
   background evolution, are constructed out of $H(z)$, it is
   practically impossible to distinguish these models from $\Lambda$CDM
   using the current data. While analyzing the observational constraints,
   we used
   supernova and  BAO data along with the information on Hubble parameter measurements.
   Our analysis
   shows that for smaller values of $\Omega_{m0}$, it is easier to distinguish amongst
   the various models. We find that tachyon and scalar field models with linear potential as
 well as scalar field model with PNGB potential are comparatively easier to distinguish
from $\Lambda$CDM. It  should be mentioned that little attention is
paid in the literature to scalar models with linear potentials
\cite{linear}; these models deserve further investigations.

   An  interesting outcome of our study is related to the small redshift range showing distinguished
   features. We find that the behavior of thawing dynamics
   around the redshift interval, $z=0.5$ $-1$,  is most sensitive to  study
   the
   deviations
   from $\Lambda$CDM. In our opinion, future observations should concentrate more on
   this particular range so as to decrease the error bars significantly.

\section{Acknowledgement}
A.A.S acknowledges the financial support provided by the University Grants Commission, Govt. Of India, through major research project grant (Grant No:33-28/2007(SR)). S.S acknowledges the financial support provided by University Grants Commission, Govt. Of India through the D.S.Kothari Post Doctoral Fellowship.

\end{document}